\documentclass[pre,twocolumn,superscriptaddress,showpacs,floatfix]{revtex4}

\usepackage{graphicx}
\usepackage{amssymb}

\begin{document}               

\def\be{\begin{equation}}
\def\ee{\end{equation}}
\def\ba{\begin{eqnarray}}
\def\ea{\end{eqnarray}}
\def\bas{\begin{eqnarray*}}
\def\eas{\end{eqnarray*}}


\title{Bouncing ball orbits and symmetry breaking effects in a three-dimensional chaotic 
billiard}
\author{B.~Dietz}
\affiliation{Institut f\"ur Kernphysik, Technische Universit\"at Darmstadt,
D-64289 Darmstadt, Germany}
\author{B.~M\"o{\ss}ner}
\affiliation{Abteilung f\"ur Angewandte Mathematik, Universit\"at Freiburg,
 D-79104 Freiburg, Germany}
\author{T.~Papenbrock}
\affiliation{Department of Physics and Astronomy, University of Tennessee,
Knoxville, TN~37996, USA}
\affiliation{Physics Division,
Oak Ridge National Laboratory, Oak Ridge, TN 37831, USA}
\author{U.~Reif}
\affiliation{Fachbereich Mathematik, Technische Universit\"at Darmstadt,
D-64289 Darmstadt, Germany}
\author{A.~Richter}
\affiliation{Institut f\"ur Kernphysik, Technische Universit\"at Darmstadt,
D-64289 Darmstadt, Germany}

\date{\today}

\begin{abstract}

We study the classical and quantum mechanics of a three-dimensional
stadium billiard. It consists of two quarter cylinders that are
rotated with respect to each other by 90 degrees, and it is classically
chaotic.  The billiard exhibits only a few families of nongeneric
periodic orbits. We introduce an analytic method for their
treatment. The length spectrum can be understood in terms of the
nongeneric and unstable periodic orbits. For unequal radii of the
quarter cylinders the level statistics agree well with predictions
from random matrix theory. For equal radii the billiard exhibits an additional
symmetry. We investigated the effects of symmetry
breaking on spectral properties. Moreover, for equal radii, we observe a small
deviation of the level statistics from random matrix theory. This led
to the discovery of stable and marginally stable orbits, which are absent for unequal 
radii.
\end{abstract}

\maketitle

\section{Introduction}

Wave chaotic phenomena are visible in a large variety of physical
systems ranging from lattice QCD~\cite{V94,VW00} to
nuclei~\cite{Brody81}, atoms~\cite{Friedrich}, mesoscopic
systems~\cite{Mirlin}, optical microcavities~\cite{Noeckel}, microwave
resonators~\cite{Stock,Sridhar,R1}, and to vibrations of macroscopic
objects~\cite{Weav,Elle}. The correspondence between the classical
dynamics and wave phenomena in the semiclassical regime is of
particular interest in such systems; for a comprehensive review, we refer the
reader to Ref.~\cite{QC,GMW}. It is best understood in
Hamiltonian systems with two degrees of freedom, whereas there is a
lack of studies of chaotic three-dimensional systems. Experimental
investigations of wave chaotic phenomena in three dimensions can be
performed with microwave resonators\cite{Deus,Wirzba,SinaiExp} and
acoustic blocks~\cite{Weav,Elle}. The resonance spectra investigated
in such experiments are described by non-scalar wave equations and are
considerably more complicated than the quantum mechanical wave
equations of a non-integrable Hamiltonian system with three degrees of
freedom.

Let us briefly review the (relatively short) list of studies of
quantum chaos in three-dimensional billiards. Prosen investigated
quantum chaotic phenomena in a three-dimensional deformed sphere
\cite{Prosen}. Primack and Smilansky~\cite{PS} unraveled the classical and
quantum mechanics of the three-dimensional Sinai billiard~\cite{Sinai},
and verified the applicability of Gutzwiller's trace
formula~\cite{Gutz,Gutzwiller}; a corresponding experimental
study of a microwave resonator was presented in
Ref.~\cite{SinaiExp}. The experimental study~\cite{Richt} of the
three-dimensional Bunimovich stadium led to a verification of the
trace formula proposed by Balian and Duplantier \cite{BaDu} for
electromagnetic systems.  A self-bound three-body system with
high-dimensional scars was studied theoretically in
Ref.~\cite{PP}. Problems involving mode coupling in three-dimensional systems
were investigated in vibrating crystals~\cite{Elle} and also in a microwave 
resonator \cite{Richt}.

The theoretical description of quantum chaos in three-dimensional
systems is in general very difficult. A full understanding of the
classical dynamics is compounded by the difficulty to visualize motion
in phase space. Moreover, the analysis of level statistics of the
corresponding quantum system requires the accurate computation of long
level sequences and can be a very time-consuming task for eigenstates
with short wave lengths. Furthermore, there are a number of open questions.
These concern, for instance, the applicability and accuracy
of semiclassical periodic orbit sums for the quantization of such
systems.  This problem was addressed by Primack and Smilansky in their
study of the three-dimensional Sinai billiard~\cite{PS}, which 
is completely chaotic. Yet, an infinite number of families of
marginally stable (bouncing ball) orbits considerably complicates the
semiclassical computation of the level density in terms of
Gutzwiller's trace formula. For the evaluation of the resonance
density in three-dimensional microwave resonators, Gutzwiller's trace 
formula does not
apply, and one has instead to use the trace formula by Balian and
Duplantier. The applicability of this periodic orbit sum was recently
investigated and confirmed for a three-dimensional stadium billiard
with chaotic dynamics \cite{Richt}. It is the purpose of the present work to study
the quantum mechanical aspects of the three-dimensional stadium
billiard further. We will focus in particular on quantum manifestations of
classical chaos, the applicability of Gutzwiller's periodic orbit sum,
and on effects related to symmetry breaking.

\begin{figure}[h]
\includegraphics[width=0.45\textwidth]{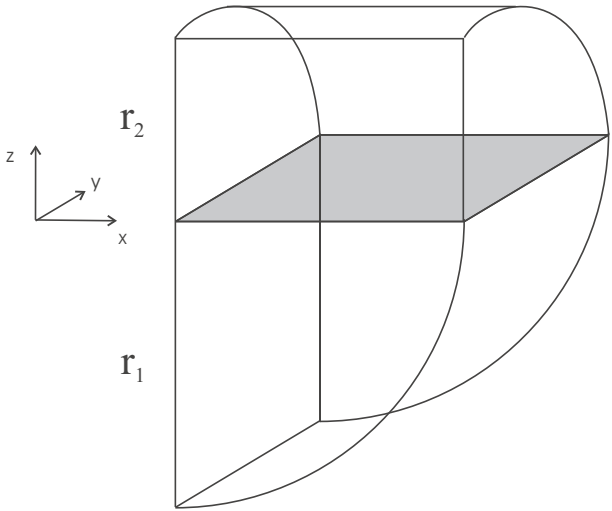}
\caption{\label{fig1}(Color online) Three-dimensional stadium billiard. The plane
$z=0$ is shaded.}
\end{figure}

The billiard depicted in Fig.~\ref{fig1} is a generalization of the
two-dimensional Bunimovich stadium~\cite{Bunimovich} to three
dimensions~\cite{Stadium3D}. It consists of two quarter cylinders that
are rotated about 90 degrees with respect to each other. For the case
that these quarter cylinders are separated by a finite distance $a$,
Bunimovich and Del Magno~\cite{Bun06} showed that this billiard is completely
hyperbolic, i.e. there is no finite measure of trajectories that is
not exponentially sensitive to changes of their initial
conditions. Earlier numerical studies suggest that the billiard of
Fig.~\ref{fig1} is also classically chaotic \cite{Stadium3D}.  In
what follows, we restrict ourselves to this case.
 
There are several reasons why the three-dimensional stadium billiard
is a promising candidate for the study of quantum chaos. First, the
classical dynamics is strongly chaotic and no stable islands are known
for $r_1\ne r_2$. 
In this work, we focus particularly on the case $r_1=\sqrt{2}r_2$
since this geometry was studied in microwave experiments \cite{Richt}
and in Ref.~\cite{Stadium3D}. The ratio was chosen irrational in order to 
avoid nongeneric quantum effects due to classical orbits of measure zero (see below).
Second, nongeneric contributions to the
level density arise due to two families of bouncing-ball
orbits. However, unlike in the case of the three-dimensional Sinai
billiard~\cite{PS}, their contribution can be computed and
subtracted. Third, for $r_1=r_2$ the billiard exhibits a high symmetry
(under reflection at the $z=0$ plane with a subsequent rotation around
the $z$-axis about $90$ degrees), and this offers the opportunity
to study symmetry-breaking effects.  Below, we will see that the level
statistics in this highly symmetric case exhibit a deviation from the
theoretical prediction for chaotic systems. This led to the discovery
of a stable and a few marginally stable orbits which were
previously unknown.

This article is organized as follows. In Sect.~\ref{nongen}, we focus
on classical periodic orbits and nongeneric modes. In Sect.~\ref{qm},
we present our analysis of the quantum mechanical spectra and discuss
level statistics and symmetry breaking.  We conclude with a summary of
the main results. Some technical aspects concerning the finite element
approximation with web-splines are deferred to the Appendix.

\section{Nongeneric modes and periodic orbits}
\label{nongen}

In this section we investigate nongeneric modes and classical periodic
orbits.  The former are an interesting, system-specific property and
must be subtracted from the staircase function before generic
properties can be analyzed; the latter are useful in a semiclassical
interpretation of quantum spectra. 

\subsection{Nongeneric modes}

We study the desymmetrized stadium billiard depicted in
Fig.~\ref{fig1}. The dynamics is limited to $x\ge 0$ and $y\ge 0$ with
specular reflections at the planes $x=0$ and $y=0$. We use $r_1\ge r_2$
and will express lengths in units of $r_2$ and wave momenta in units
of $r_2^{-1}$. For $r_1\ne r_2$, the billiard is fully desymmetrized;
for $r_1=r_2$, it is still symmetric under reflection at the plane
$z=0$ and a subsequent rotation by $\pi/2$ around the $z$-axis. We
label eigenstates by their wave momentum $k$.  Employing a recently
developed method which is based on finite elements and seeks solutions
of the Schr\"odinger equation with Dirichlet boundary conditions, we
computed the lowest 1200 levels up to $k r_2\approx 35$.  More details
on this method are given in the appendix.

\begin{figure}[h]
\includegraphics[width=0.45\textwidth]{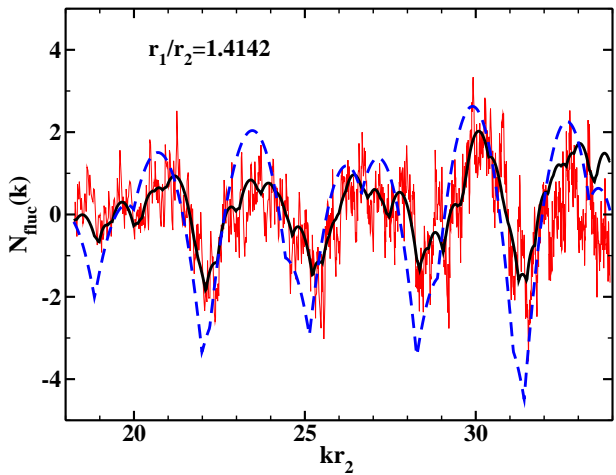}
\caption{\label{fig2}(Color online) Fluctuating part $N_{\rm fluc}$ of the
staircase function for the geometry with $r_1/r_2=\sqrt{2}$ (thin
line). Contributions from bouncing-ball orbits (dashed line) and from
all nongeneric modes (thick line).}
\end{figure}

The staircase function 
\be 
N(k) = \sum_{n=1}^X \Theta (k-k_n)  
\ee
with $\Theta(x)$ denoting the unit step function counts the levels below 
a given $k$. It is
the sum of a smooth part $N_{\rm smooth}(k)$ and a fluctuating part
$N_{\rm fluc}(k)$, i.e., $N(k) = N_{\rm smooth}(k) + N_{\rm
fluc}(k)$. The smooth part is given by the Weyl formula which is a
polynomial of degree three in $k$, and is subtracted from
$N(k)$. Figure~\ref{fig2} shows the remaining fluctuating part.  One
identifies large-scale oscillations that grow in amplitude with
increasing wave momentum. These fluctuations are due to the nongeneric
modes of the billiard associated with the bouncing-ball orbits
perpendicular to the flat boundaries of the billiard and the orbits
within the $z=0$ plane.

\begin{figure}[h]
\includegraphics[width=0.45\textwidth]{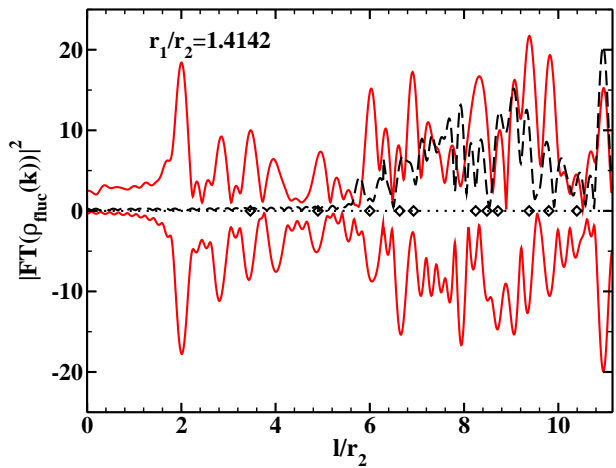}
\caption{\label{fig3}(Color online) Upper part: Length spectrum of the
three-dimensional stadium billiard for $r_1/r_2=\sqrt{2}$ (full
line). Contributions from unstable periodic orbits (dashed
line). Lower part: Sum of contributions from nongeneric modes and 
unstable periodic orbits (full line). Peaks at multiples of $2r_1$ and $2r_2$ 
correspond to the bouncing-ball orbits in the two quarter cylinders. Diamonds: 
Lengths of nongeneric orbits in the $z=0$ plane.}
\end{figure}

The fluctuating part of the level density is obtained by
differentiation, i.e.  $\rho_{\rm fluc}(k)={d\over dk}N_{\rm
fluc}(k)$. The absolute value squared of the Fourier transform of this
quantity yields the length spectrum. It is shown in the upper part of Fig.~\ref{fig3}
for $r_1=\sqrt{2}r_2$. The peaks in the length spectrum appear at the
lengths of classical periodic orbits~\cite{Gutzwiller}. The most prominent
peaks are at lengths that are multiples of the bouncing-ball orbits
with lengths $2r_1$ and $2r_2$, respectively. There are other peaks
that can be identified with the lengths of orbits inside the
rectangular $z=0$ plane. Both types of orbits are nongeneric due to
their particular stability properties.  As we are interested in
generic properties, we have to extract these contributions.

First, let us consider the bouncing-ball orbits.  There are two
families of bouncing-ball orbits which bounce back and forth between
the two parallel planes of distance $r_1$ and $r_2$ in the two quarter
cylinders, respectively.  These orbits are marginally stable as they
have zero Lyapunov exponents. According to semiclassical periodic
orbit theory, they are associated with long-range oscillations of the
quantum mechanical density of states.

We follow Ref.~\cite{Wirzba} for the computation of the contribution
of bouncing-ball orbits to the staircase function, and focus on the
bouncing-ball modes in the upper quarter cylinder first. The number of
bouncing-ball (bb) modes up to momentum $k$ is given by
\ba
\label{Nbb}
N_{\rm bb}(k) &=& {\rm Tr}_{{\rm bb}}\Theta\left( {\hbar^2k^2\over 2m}- 
\hat{H}\right)\nonumber\\ 
&=& \int {dz dk_z\over 2\pi} \int {dy dk_y\over 2\pi} 
\sum_{\mu=1}^\infty \nonumber\\
&&\Theta\left(k^2 - k_z^2 -k_y^2- k_{x,\mu}^2 \right)
\nonumber\\ 
&+& \left( (x,y)\to (y,x)\right) \ .
\ea
Here, 
\be
\label{kmode}
k_{x,\mu}\equiv {\mu\pi\over l_x}\ ,
\ee
and $l_x=r_1$, $l_y=r_2$. This ansatz for the number of bouncing-ball modes is
intuitively clear. The trace is performed in the semiclassical
approximation as a phase space integral in the directions
perpendicular to the bouncing-ball orbits, while these modes are
quantized along the orbit. The integrations can be performed, and one
obtains~\cite{Wirzba}
\ba
\label{Nbbfin}
N_{\rm bb}(k) &=& {r_1^2\over 16}\sum_{\mu\ge 1} 
\left(k^2 - k_{2,\mu}^2\right) \Theta\left(k^2 - k_{2,\mu}^2\right) \nonumber\\
&+& 
{r_2^2\over 16}\sum_{\mu\ge 1} \left(k^2 - k_{1,\mu}^2\right) 
\Theta\left(k^2 - k_{1,\mu}^2\right) \ .
\ea
To obtain the fluctuating contribution $N_{\rm bb, fluc}$, we
subtract a polynomial of third degree in $k$ from the
expression~(\ref{Nbbfin}). We confirmed that the leading coefficients
of this polynomial agree with the coefficients of the Weyl formula. An
analytical formula is presented in Ref.~\cite{Wirzba}.  As evident from
Fig.~\ref{fig2}, the bouncing-ball orbits (dashed line) describe the gross
oscillations of the fluctuating staircase function rather well. To
account for finer details, we also have to consider nongeneric
modes inside the $z=0$ plane.

The rectangular $z=0$ plane is special, since the classical motion is
regular inside this plane but unstable with respect to deviations out
of this plane. Note that this does not spoil the hyperbolicity of the
billiard since the set of orbits restricted to the $z=0$ plane are of
measure zero in classical phase space. However, remnants of families
of classical orbits inside this plane can also be associated with
long-range fluctuations in the quantum mechanical level density for the
following two reasons. First, the periodic orbits inside this plane
come in two-dimensional families opposed to the isolated periodic
orbits in hyperbolic systems. Second, these orbits are less unstable
than most truly three-dimensional periodic orbits and thus enter any
periodic-orbit sum with a greater weight. It is thus necessary to include 
the nongeneric modes associated with the $z=0$ plane.  

Extending the results from Ref.~\cite{Wirzba}, the number of
modes associated with nongeneric orbits is given by the quantum
mechanical trace over nongeneric (ng) modes 
\ba
\label{Nng}
N_{\rm ng}(k) &=& {\rm Tr}_{{\rm ng}}\Theta\left( {\hbar^2k^2\over 2m}- 
\hat{H}\right)\\ 
&=& \int {dz dk_z\over 2\pi} 
\sum_{\mu,\nu=1}^\infty 
\Theta\left(k^2 - k_z^2 - k_{x,\mu}^2 - k_{y,\nu}^2\right) \nonumber \ .
\ea

This ansatz is again transparent: the trace is approximated by a
phase-space integral in the $z$-direction, while the rectangular
sections (with $z$-dependent area) perpendicular to the $z$-axis are
explicitly quantized. We thus quantize the motion inside this plane
adiabatically, quite similar to the approach that Bai {\it et
al.}~\cite{Taylor} applied to the Bunimovich stadium.  Note that
Eq.~(\ref{Nng}) accounts both for the bouncing-ball orbits and for the
modes associated with the $z=0$ plane.

In Eq.~(\ref{Nng}), the wave momenta $k_{x,\mu}$
and $k_{y,\nu}$ are defined as in Eq.~(\ref{kmode}) but with the
$z$-dependent radii $l_x$ and $l_y$. We have
\ba
l_x(z) &=& \left\{
     \begin{array}{ll}
     r_1 & \mbox{for $z<0$}, \\
     \sqrt{r_1^2-z^2} & \mbox{for $z\ge 0$}.
     \end{array}
     \right. \nonumber\\
l_y(z) &=& \left\{
     \begin{array}{ll}
     r_2 & \mbox{for $z\ge 0$}, \\
     \sqrt{r_2^2-z^2} & \mbox{for $z<0$}.
     \end{array}
     \right.
\ea
The integrations in Eq.~(\ref{Nng}) can be performed
analytically, and one obtains
\be
\label{Nng_fin}
N_{\rm ng}(k) = {1\over \pi} {\sum_{\mu,\nu\geq 1}^{}}^\prime 
\left( A(\kappa_{\mu\nu},r_1,r_2) 
+ A(\kappa_{\mu\nu},r_2,r_1)\right) \ .
\ee
Here
\ba
\label{A}
A(\kappa_{\mu\nu},r_1,r_2)&\equiv& 
r_1\sqrt{\kappa^2_{\mu\nu} - k_{2,\nu}^2}\nonumber\\
&&\left( E(\kappa_{\mu\nu})-(1-\kappa^2_{\mu\nu}) K(\kappa_{\mu\nu})\right)\  
\ea
with
\be
\label{kappa}
\kappa_{\mu\nu}=\kappa_{\mu\nu}(k,r_1,r_2)\equiv\left({k^2-k_{1,\mu}^2
  -k_{2,\nu}^2 \over k^2-k_{2,\nu}^2}\right)^{1/2} \   
\ee 
contains the dependence on the wave momentum and the radii of the
billiard.  The wave momenta $k_{1,\mu}$ and $k_{2,\nu}$ are defined in
Eq.~(\ref{kmode}).  In Eq.~(\ref{A}), $K(\kappa)$ and $E(\kappa)$ are
the complete elliptical integrals of the first and second kind,
respectively. The prime in the sum of Eq.~(\ref{Nng_fin}) indicates
that the summation is limited to values of $\mu,\nu$ such that the
square roots in Eqs.~(\ref{A}) and (\ref{kappa}) are real.

The smooth part of the staircase function is again a polynomial of
degree three and is subtracted. The fluctuating part of $N_{\rm
ng}(k)$ is shown in Fig.~\ref{fig2} as a thick line. Clearly, the gross
and also finer oscillations are accounted for. This enables us to analyze
the length spectra depicted in Fig.~\ref{fig3}. The upper part shows
the length spectra, i.e. the Fourier transform of the fluctuating part
of the density of states, the lower part that of the  nongeneric modes 
and the unstable periodic orbits, which 
has been computed based on Eq.~(\ref{Nng_fin}) and the Gutzwiller 
trace formula (see next section). The peaks at multiples of the lengths
$2r_1$ and $2r_2$ are due to the bouncing-ball modes. The peaks marked
by a diamond are due to the nongeneric orbits in the $z=0$ plane. The
remaining peaks must thus be associated with lengths of isolated
periodic orbits, and we turn to their analysis in the following
subsection.

\subsection{Unstable periodic orbits}

Following Gutzwiller's periodic orbit theory~\cite{Gutzwiller}, the
semiclassical approximation of the quantum mechanical density of
states is given in terms of the periodic orbits of the underlying
classical system. This sum is infinite such that approximations have
to be invoked. Here, we are interested in the length spectrum,
i.e. the power spectrum of the Fourier transform of the fluctuating part 
of the level
density. The length spectrum up to length $l$ is given in terms of all
periodic orbits up to length $l$, and this is a finite (but usually
with $l$ exponentially increasing) number of periodic orbits.

The search for periodic orbits is a cumbersome task. Here, we use two
different methods and focus on periodic orbits outside the $z=0$
plane. The first method considers the Poincar{\'e} surface of section
(PSOS) defined by $z=0, p_z>0$, and constructs the PSOS map. Periodic
orbits are fixed points of this map. They are found by starting a
large number of trajectories in the PSOS and by using a Newton-Raphson
algorithm to find fixed points in the vicinity of each such
trajectory.

The second method is based on a symbolic code and utilizes the fact
that the length of a periodic orbit is a local extremum under
variation of the points of reflection along the billiard's
boundary. This procedure is similar to the one described in
Ref.~\cite{PS}. We consider an open billiard system which is the
infinite periodic extension of our billiard, and assign the letters
``+'' and ``-'' to reflections on the curved parts of the upper and
lower quarter cylinder, respectively. Periodic orbits outside the
$z=0$ plane can certainly be described as words composed from these
two letters. We have no proof that there is a one-to-one
correspondence between this symbolic code and the periodic orbits of
our system. Given a word from this alphabet, we construct a random
closed orbit as follows. For each ``+'' (``-'') letter of the word, we
choose a random point on the curved surface of the upper (lower)
quarter cylinder. We connect this sequence of points by straight lines
and compute the length of this closed orbit. Then we vary the
positions of the random points in order to find a local minimum of the
length.

Both methods yield a considerable number of periodic orbits, and we
consider the union of the resulting sets as our set of periodic orbits. Once a
periodic orbit is found, we compute its length, monodromy matrix and its
Maslov index following Ref.~\cite{Sieber}. Recall that the monodromy
matrix is the tangent map and encodes the stability properties of the
periodic orbit. For unstable orbits, its eigenvalues come in real
pairs $\Lambda_+, \Lambda_-$ with $\Lambda_+\Lambda_-=1$ or in complex
quadruples $\Lambda_+, \Lambda_-,\Lambda_+^*, \Lambda_-^*$ with
$\Lambda_+\Lambda_-=1$. Stable orbits have two pairs of complex
eigenvalues with modulus one.

We performed our most extensive search for the billiard with the geometry
$r_1=\sqrt{2}r_2$ and determined more than 2000 periodic orbits. We
believe that our list is fairly complete for the shortest orbits. All
periodic obits we found were unstable. This extends and confirms the
numerical results of Ref.~\cite{Stadium3D} and strongly suggests that
the billiard is completely chaotic.  The quantum mechanical results
presented below support this picture. 
The contribution of the unstable periodic orbits to the length
spectrum is shown in the upper part of Fig.~\ref{fig3} as a dashed
line.  Clearly, unstable periodic orbits contribute significantly to
the length spectrum, particularly for length $l/r_2\gtrsim 6$ where the
contributions from nongeneric modes become less dominant.

However, for the case $r_1=r_2$ we found a stable periodic orbit. This
was unexpected, since earlier studies~\cite{Stadium3D} yielded a
positive Lyapunov exponent for a sample of $10^4$ randomly chosen
trajectories, and not one of the trajectories was found to be
stable. The shortest stable periodic orbit we found has a length
$l/r_2\approx 10.1706$. Note that the eigenvalues of the monodromy matrix 
corresponding to this orbit are on the unit circle and very close to the real axis.
Thus it is difficult to distinguish this orbit from a marginally stable orbit.
We also obtained periodic orbits (in addition to
the bouncing-ball orbits) that are marginally stable, i.e. the
eigenvalues of the monodromy matrix are real with modulus one. The
shortest of these orbits has a length $l/r_2\approx 5.4981$. Other
marginally stable orbits have periods $l/r_2\approx 5.5153, 6.2701,
6.4738$. 

\begin{figure}[h]
\includegraphics[width=0.45\textwidth]{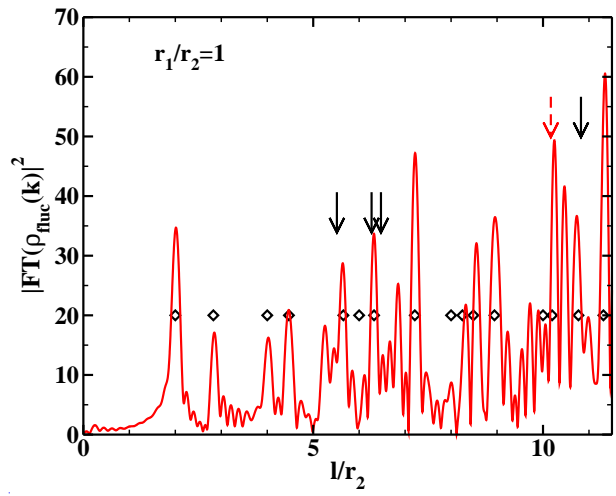}
\caption{\label{fig4}(Color online) Full line: Length spectrum for
$r_1=r_2$. Diamonds: Length of bouncing-ball orbits and of nongeneric
orbits in the $z=0$ plane. Full arrows: Length of marginally stable orbits. Dashed arrow: length of stable periodic orbit.}
\end{figure}

To study the relevance of these orbits, we show the length spectrum of
the billiard for $r_1=r_2$ in Fig.~\ref{fig4}. Diamonds denote the
lengths of bouncing-ball orbits and of nongeneric periodic orbits
associated with the $z=0$ plane; they account for many peaks in the
length spectrum. We also computed the lengths, monodromy matrices and Maslov
indices of the 400 shortest unstable and marginally stable orbits by 
proceeding as in the case $r_1=\sqrt{2}r_2$. However, we have no semiclassical 
theory for the computation of the density of states for the marginally stable
and the stable orbit, and we are not able to disentangle their contribution to the
length spectrum from that of the nongeneric modes. For this reason we show in Fig.~\ref{fig4}
the length spectrum of the billiard and indicate the lengths of the nongeneric orbits 
by diamonds, and those of the marginally stable orbits by the full arrows.
The dashed arrow marks the length 
of the stable orbit. It is difficult to clearly identify its impact 
on the length spectrum, since the peak around length
$l/r_2\approx 10.19$ can be attributed to the stable orbit and/or to a
nongeneric orbit. Recall that a stable orbit leaves a strong imprint
in the length spectrum if the stable island around it is sufficiently
large (in units of $(2\pi\hbar)^3$) to accommodate
eigenstates. To obtain an estimate for
the phase-space volume of the elliptical island we started bundles of 
$21^4$ trajectories in the PSOS close to the stable periodic orbit. All 
trajectories departed from the stable orbit after a few intersections 
with the PSOS. Within the achievable numerical accuracy we may thus conclude, 
that either the volume of the phase space associated with the stable orbit is 
very small or that contrary to our numerical results it is only marginally stable. 
Note that there are several marginally stable orbits
associated with the third arrow from the left. These orbits have
almost identical lengths, and the visual inspection indicates that
they are whispering gallery orbits. In this case, interference effects
might explain why there is no clear peak in the length spectrum
associated with these orbits. A similar effect has been found in the
two-dimensional stadium billiard~\cite{Sie}. In the next section we
present evidence that the stable and marginally stable orbits explain
peculiarities in the level statistics of the billiard with $r_1=r_2$.

\section{Spectral properties and symmetry breaking}
\label{qm}

\begin{figure}[h]
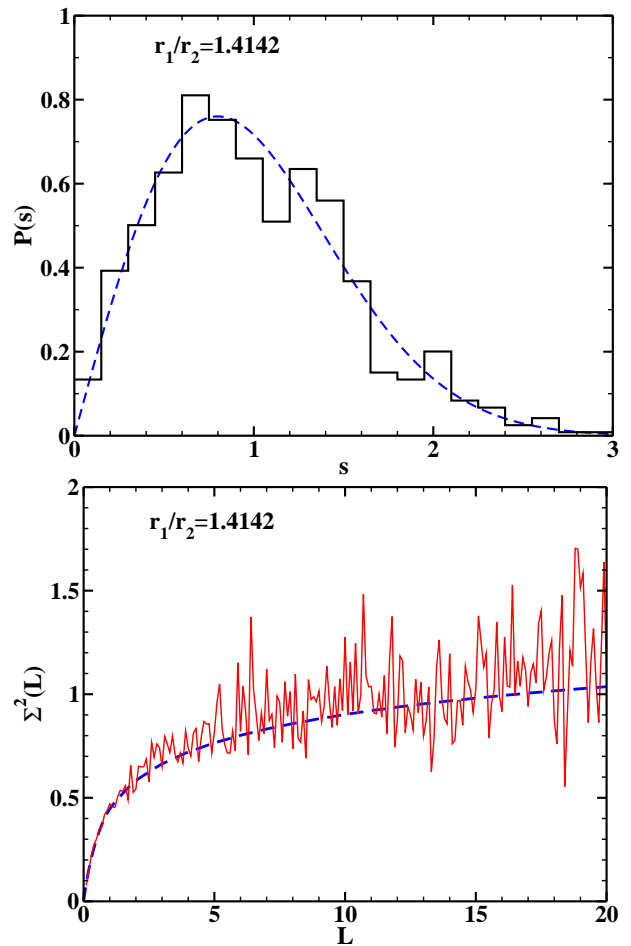

\includegraphics[width=0.45\textwidth]{fig5a.eps}
\includegraphics[width=0.45\textwidth]{fig5bneu.eps}
\caption{\label{fig5}(Color online) Upper panel: Nearest neighbor
spacing distribution (histogram) of the billiard for $r_1=\sqrt{2}r_2$
compared to the GOE prediction (dashed line). Lower panel:
$\Sigma^2$-statistics (thin line) compared to the GOE prediction
(dashed line).}
\end{figure}

In this section we investigate statistical properties of the
eigenvalues of the three-dimensional stadium billiard for a varying
ratio of the radii $r_1/r_2$ of the two quarter cylinders, 
where the volume of the billiard is kept fixed. 
We computed the lowest 1200 levels with a recently developed finite element
method~\cite{Moessner} described in the appendix. The 
lowest 200 levels are discarded since we are interested in the 
correspondence between classical and quantum mechanics.
In order to get rid of system-dependent properties,
we need to rescale the eigenvalues to unit mean spacing and to extract
the contributions of the nongeneric modes to the fluctuating part of
the resonance density. This is done by replacing the computed wave
numbers $k$ by $\tilde k=N_{\rm smooth} (k)+N_{\rm ng, fluc}(k)$
(see \cite{R1,Wirzba}). As the underlying classical dynamics is
completely chaotic, we expect agreement of the statistical properties
of the unfolded eigenvalues with those of random matrices drawn
from the Gaussian orthogonal ensemble (GOE) if $r_1$ is chosen not
equal to $r_2$ \cite{bgs}. In Fig.~\ref{fig5} we show the nearest-neighbor
spacing distribution and the $\Sigma^2$-statistics for a ratio of
radii $r_1/r_2=\sqrt{2}$, i.e. for the case considered in the
experiments with the microwave cavity of the shape of a
three-dimensional stadium billiard \cite{Richt}. Both curves agree
well with the corresponding ones for random matrices from the
GOE. This shows that first the three-dimensional quantum stadium
billiard behaves like a generic quantum system with chaotic classical
dynamics, and second that our procedure of extracting the contribution
of nongeneric modes is applicable and complete.

We recall that the experiment~\cite{Richt} with a microwave
resonator of the shape of a three-dimensional stadium 
billiard with $r_1/r_2=\sqrt{2}$ revealed deviations of the spectral 
properties from GOE-behavior. This was attributed to a partial decoupling
between electromagnetic TE and TM modes in the low-frequency
regime. Note, however, that this system differs from the quantum
billiard considered in the present work due to the vectorial character
of the underlying wave equations. As a consequence, the trace formula
by Balian and Duplantier~\cite{BaDu} has to be employed for the semiclassical 
computation of the length spectrum. It differs from Gutzwiller's trace formula 
in the occurrence of a factor associated
with the polarization, and as a consequence, only periodic orbits with
an even number of reflections enter. The evaluation of this factor 
showed that the electric and magnetic modes are decoupled on a considerable 
fraction of the short periodic orbits. This finding supports the 
interpretation of the observed deviations from GOE-behavior in terms of
a partial decoupling between the TE and the TM modes. In \cite{Richt}  
the experimental length spectrum was well reproduced by the theoretical
calculations based on the trace formula derived by Balian and Duplantier. 
As there the polarization of the electric 
field is taken into account for each unstable periodic orbit individually, 
this good agreement is not in contradiction to the
discrepancy obtained for the spectral properties.

\subsection{Symmetry breaking}

For $r_1=r_2$, the billiard is symmetric with respect to reflections at
the $z=0$ plane and a subsequent $\pi/2$-rotation about the
$z$-axis. Accordingly, in this case the wave functions of the billiard
are symmetric or antisymmetric under the symmetry operation and belong
to different irreducible representations (IR). Assuming that the
billiard is chaotic, the spectral properties of levels within each IR
are expected to coincide with those of random matrices from the GOE.
However, the spectral statistics for the whole set of eigenvalues
should coincide with RMT for a superposition of
two independent GOEs.

\begin{figure}[h]
\includegraphics[width=0.45\textwidth]{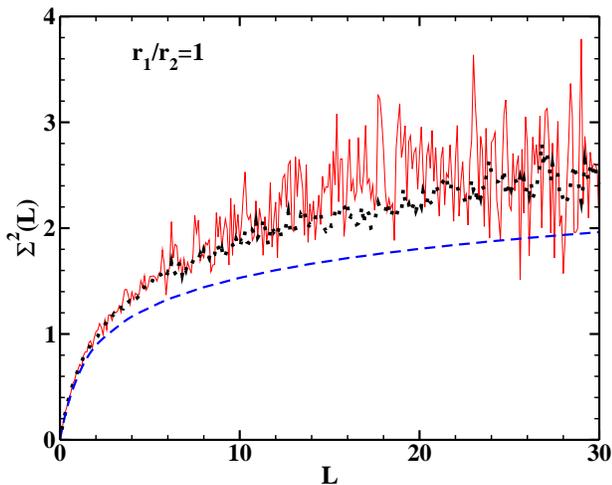}
\caption{\label{fig6}(Color online) $\Sigma^2$-statistics for the
billiard with $r_1=r_2$ (thin line) compared to the GOE prediction
(dashed line) and to a random matrix model which also includes a
Poissonian sequence of random levels (dots).}
\end{figure}

\begin{figure}[h]
\includegraphics[width=0.45\textwidth]{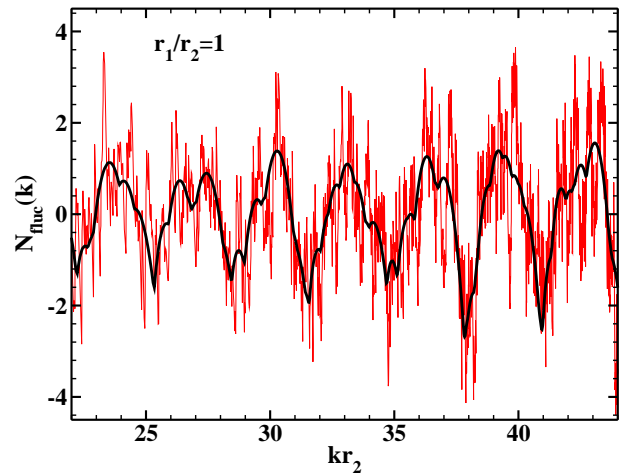}
\caption{\label{fig7}(Color online) The fluctuating part $N_{fluc}$ of
the staircase function for the billiard with $r_1=r_2$ (thin line)
compared to the contributions from the nongeneric modes (thick line).}
\end{figure}

Let $H_1$ and $H_2$ be two random matrices drawn from the GOE and consider
an ensemble of random matrices of the form \cite{GMW,Rosen,Leitner}
\begin{equation}
H=\pmatrix{H_1 & 0   \cr
           0   & H_2 \cr
          }\, .
\label{HamMod}
\end{equation}
For a generic chaotic system with two symmetry classes, the spectral
properties of the eigenvalues associated with each symmetry class are
given by those of the random matrices $H_1$ and $H_2$,
respectively, whereas the complete set of eigenvalues is described by
those of $H$ itself. Since, in our case, the number of eigenvalues
associated with the two symmetry classes are (approximately) equal,
the matrices $H_1$ and $H_2$ are chosen of equal dimension for the
theoretical description of the spectral properties of the quantum
billiard. In Fig.~\ref{fig6} we compare the $\Sigma^2$-statistics for
the eigenvalues of the quantum billiard (thin line) with that of an
ensemble of random matrices of the form Eq.~(\ref{HamMod}) (dashed
line), which is known analytically \cite{Leitner}.  We observe
significant deviations, which cannot be explained by an additional
family of nongeneric modes. Indeed, Fig.~\ref{fig7} shows the
fluctuating part of the staircase function $N_{\rm fluc}(f)$ for the
case $r_1=r_2$ (thin line) and the contributions of the nongeneric modes resulting
from Eq.~(\ref{Nng_fin}) (thick line). As for the case $r_1/r_2=\sqrt{2}$, the
smooth oscillations of $N_{\rm fluc}(f)$ are well described by our
expression. From this we may conclude that the adiabatic method
described in Section \ref{nongen} yields a good approximation for the
contribution of the nongeneric modes to the staircase function also
for $r_1=r_2$. We also verified, that the deviations are not due to
insufficient numerical accuracy in the computation of the eigenvalues.
How can this puzzle be understood?

Recall that the billiard has a stable orbit for $r_1=r_2$ and several
marginally stable orbits besides the bouncing ball orbits.
This suggests that their presence causes the deviation depicted in
Fig.~\ref{fig6}. In order to test this hypothesis, we considered the
following random matrix ensemble. Each random matrix consists of two
block matrices $H_1$ and $H_2$ of equal dimension that are drawn from
the GOE. These model the two symmetry classes of eigenstates
associated with the chaotic part of the billiard (see
Eq.~(\ref{HamMod})). An additional diagonal random matrix of much
smaller dimension models eigenstates associated with the few stable
orbits. This diagonal matrix thus exhibits Poisson statistics. The
dimension of the latter matrix was chosen equal to 25, that of $H_1$
and $H_2$ equal to 250. The $\Sigma^2$-statistics of this random
matrix model is plotted as a dotted line in Fig~\ref{fig6}. It agrees
very well with that of the billiard. This suggests that the stable and
marginally stable orbits are responsible for the observed deviations
from the prediction of standard RMT. Most interestingly, the stable
orbits disappear immediately when the ratio $r_1/r_2$ differs slightly
from one.

For geometries $r_1\ne r_2$, the symmetry of the billiard is broken.
The underlying quantum system ``sees'' this symmetry breaking once the
perturbation due to the symmetry breaking is of the order of the mean
level spacing, or, alternatively, once the geometric distortions due
to the symmetry breaking are of the scale of one wave length.  For a
completely broken symmetry, RMT predicts agreement of the spectral
properties of the billiard with those of one GOE.  Quantum
mechanically, however, we found that for $r_1\approx r_2$, the
spectral properties coincide with those of random matrices applicable
to chaotic systems with a partially broken symmetry
~\cite{GMW,Rosen,Leitner,Alt,SymBrech},

\begin{equation}
H=\pmatrix{H_1 & 0   \cr
           0   & H_2 \cr
          }\, +
\sqrt{\lambda} D\pmatrix{0   & V   \cr
         V^T & 0   \cr
          }\, \ .
\label{HamMod1}
\end{equation}
In this random matrix model, the first (second) matrix preserves
(breaks) the symmetry. The size of the symmetry breaking is set by 
the dimensionless parameter
$\lambda$ measured on the scale of the mean spacing $D$.  Here as in
Eq.~(\ref{HamMod}), $H_1$ and $H_2$ are random matrices drawn from the
GOE.  The symmetry breaking is modeled by
the off--diagonal blocks $V$ and $V^T$, where the random matrix $V$ is
real with no symmetries. For this random matrix model, the
$\Sigma^2$-statistics is known analytically for arbitrary values of
$\lambda$.

\begin{figure}[h]
\includegraphics[width=0.45\textwidth]{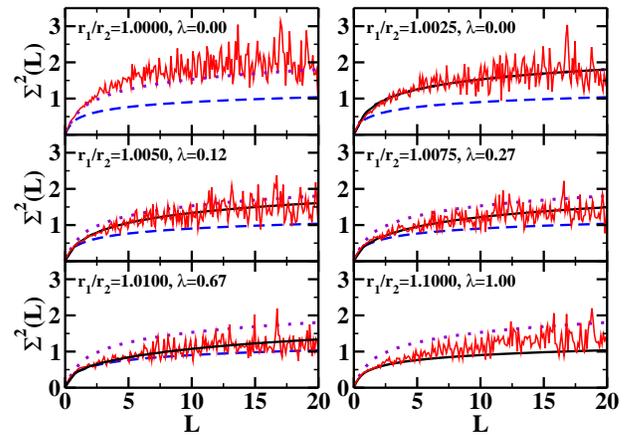}
\caption{\label{fig8}(Color online) $\Sigma^2$-statistics for
various ratios $r_1/r_2$ of the billiard (thin line) compared to that
of two GOEs (dotted line), one GOE (dashed line), and for the random matrix
model in Eq.~(\ref{HamMod}) (thick line).}
\end{figure}

Figure~\ref{fig8} shows the $\Sigma^2$-statistics (thin line) for an
increasing ratio $r_1/r_2$ of the radii of the billiard. For
comparison, we also show the $\Sigma^2$-statistics for one GOE (dashed
line), for two GOEs (dotted line), and for the random matrix
model~(\ref{HamMod1}) (thick line). The values of $\lambda$ given in the 
figure are obtained by a fit of the model~(\ref{HamMod1}) to the data.
For $r_1/r_2=1.0025$ there is
perfect agreement with the random matrix model Eq.~(\ref{HamMod}) that
describes chaotic systems with a conserved symmetry. On the one hand,
the ratio $r_1/r_2=1.0025$ deviates (sufficiently strong) from one and
the stable island has disappeared. On the other hand, this ratio is
still so close to one that the quantum mechanics is unable to resolve
the symmetry breaking. For increasing values of the ratio $r_1/r_2$,
the symmetry breaking is revealed in the spectral statistics, and the
parameter $\lambda$ in Eq.~(\ref{HamMod1}) increases from zero.
Eventually, for $r_1/r_2\approx 1.1\ldots 1.2$, the $\Sigma^2$-statistics 
approaches that of one GOE. 
This result is in agreement with a semiclassical estimate. The symmetry breaking
is resolved for wave lengths $2\pi/k \lesssim |r_1-r_2|$. We have
maximal wave momenta $kr_2\approx 35$ and can thus resolve symmetry
breaking of the order $r_1/r_2\gtrsim 1+ 2\pi/(kr_2)\approx
1.18$. Note that we can also base our semiclassical estimate on the
smooth part $N_{\rm smooth}(k)$ of the staircase function. This
function is quadratic in the symmetry-breaking parameter
$(r_1-r_2)$. At maximal wave momenta $kr_2\approx 35$ nonzero values of 
this parameter lead to significant changes of $N_{\rm fluc}(k)$
for $r_1/r_2\gtrsim 1.18$.  In conclusion, the classically abrupt change
of the symmetry properties of the system is accompanied by a gradual
change of the spectral properties of the corresponding quantum system
from those of chaotic systems consisting of a superposition of two
symmetry classes to those of chaotic systems with no further
symmetries.

We add here some more notes concerning the experiment with a microwave
resonator~\cite{Richt}. There, the spectral properties were also described with
the model given in Eq.~(\ref{HamMod1}), where one of the block
matrices $H_1,H_2$ depicts the properties of the TE, the other those
of the TM modes and deviations from GOE-behavior were interpreted as
due to a partial decoupling of them. In the experiment the ratio
$r_1/r_2\simeq\sqrt{2}$ with $r_1=200.0$~mm and $r_2=141.4$~mm was kept
fixed, while the value of the resonance frequency $f$, that is of
$k=\frac{2\pi f}{c}$ with $c$ the velocity of light, was varied up to
$f=20$~GHz.

Is for this choice of the radii and of the frequency range the 
symmetry breaking discussed above observable? 
The breaking of the symmetry existent for $r_1=r_2$ is resolvable 
for wavelengths smaller than 
$\vert r_1-r_2\vert$, that is for $2\pi/k=c/f \lesssim |r_1-r_2|=58.6$~mm. 
Accordingly, for excitation frequencies $f\gtrsim 5.12$~GHz
good agreement of the spectral
properties with those of random matrices from the GOE are
expected. In the experiment, however, deviations from one GOE
were observed up to approximately 17~GHz. Hence, they cannot be explained with 
the particular mechanism of symmetry breaking discussed above. In order to 
resolve the discrepancy between theory and experiment numerical computations 
of the full vectorial Helmholtz equation are desired.

\subsection{High-lying states}

Strictly speaking, the semiclassical approximation and the
Bohigas-Giannoni-Schmit~\cite{QC,BGS,heusler} conjecture,
which states, that the spectral fluctuation properties in quantum systems
with a chaotic classical dynamics coincide with those of random matrices drawn
from the Gaussian random-matrix ensembles, apply in the regime when
the wave length is the smallest length scale of the system,
i.e. $kr_2\gg 1$ must hold. Nevertheless, we saw in the previous
section that the semiclassical analysis of the length spectrum is also
useful for the low-lying states. In this subsection, we compute
high-lying quantum states of the billiard and perform level
statistics. For the computation of high-lying levels, we employ
Prosen's generalization~\cite{Prosen} of the two-dimensional method by
Vergini and Saraceno~\cite{Vergini} to three-dimensional
billiards. This method is particularly suited for high-lying
eigenstates and determines a stretch of eigenstates around some
arbitrary wave momentum $k_0$ with $k_0 r_1\gg 1$. This method yields
1669 levels in the regime $79\lesssim k r_2 \lesssim 86$.

\begin{figure}[h]
\includegraphics[width=0.45\textwidth]{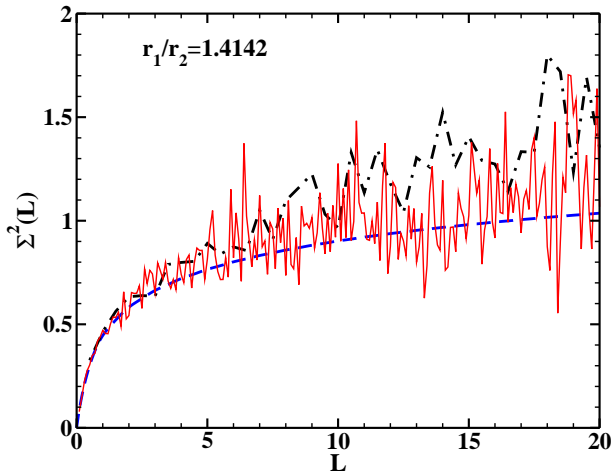}
\caption{\label{fig9}(Color online) $\Sigma^2$-statistics of the
high-lying stretch of levels around $kr_2\approx 80$ (dashed-dotted
line) compared to that of the low-lying levels (thin line) and the GOE
prediction (dashed line).}
\end{figure}

Figure~\ref{fig9} compares the spectral properties of the eigenvalues
in the lowest part of the spectrum (thin line) with those of the
high-lying eigenvalues (dashed line) and the GOE (dashed-dotted line).
While the $\Sigma^2$-statistics of the low-lying set of eigenvalues
agrees well with that of random matrices from the GOE, we observe
significant deviations for that of the high-lying ones. This points to
inadequacies of our procedure to extract the contribution of the
nongeneric modes to the fluctuating part of the resonance density.
Recall that our semiclassical formula~(\ref{Nng_fin}) takes account of
the leading nongeneric modes. These are contributions that scale as
$k^3$ with increasing wave momentum. Next-to-leading order
contributions scale as $k^2$. These are only partly included. In our
ansatz~(\ref{Nng}) we integrate over rectangles with a $z$-dependent
area. There are many billiards of different shape (but identical
volume) that have rectangular cross sections. However, evidently, our
procedure is well suited for small or moderate values of $k r_2$, but
it seems to be insufficient in the semiclassical regime.

\section{Summary}
We studied the classical and quantum mechanics of the
three-dimensional stadium billiard. This billiard consists of two
quarter cylinders that are rotated by 90 degrees with respect to each
other and is classically chaotic for unequal radii of the quarter
cylinders. We studied the nongeneric modes of the billiard that are
due to bouncing-ball orbits and orbits within a rectangular cross
section of the billiard, and gave analytical expressions for their
contribution to the staircase function. The quantum mechanical length
spectrum can be understood in terms of the nongeneric modes and
unstable periodic orbits. For unequal radii of the two quarter
cylinders, the level statistics agrees with predictions from random
matrix theory. For equal radii, we found deviations from standard
random matrix theory and these can be attributed to a stable and 
a few marginally stable
orbits. We modeled a random matrix ensemble on this basis and found
very good agreement with our data. We studied the level statistics as
a function of the ratio of the two radii. For slightly unequal radii,
there is a sudden transition from a system with mixed phase space to a
system with a chaotic phase space. For larger differences of the radii,
we follow the symmetry-breaking transition towards the spectral
statistics of one GOE.

\section*{APPENDIX. FINITE ELEMENT METHOD}
\label{finele}

In order to calculate a complete sequence of eigenvalues
for the three-dimensional stadium billiard,
we use a finite element method 
(see for example \cite{Hackbusch}, ch. 11).
This discretization of the eigenvalue problem
\begin{equation}
        \label{helmholz}
        \begin{array}{rcll}
                -\Delta \psi_i & = & k_i^2 \psi_i & \mbox{ in } \Omega,\\
                        \psi_i & = & 0            & \mbox{ on }
                                                    \partial \Omega,
        \end{array}
\end{equation}
leads to a generalized eigenvalue problem
\begin{equation}
        \label{geneig}
        A^h \psi_i^h
        =
        \lambda_i^h M^h \psi_i^h,
\end{equation}
where $A^h,M^h \in \mathbb R^{n \times n}$ are finite dimensional
matrices. The eigenvalues $\lambda^h_i$ and eigenfunctions $\psi^h_i$
of the discrete problem approximate the eigenvalues/functions of
the continuous problem (\ref{helmholz}). Here the parameter
$h\in\mathbb R^+$ represents the fineness of the test functions used
in the discretization process.
With the usual assumptions (see \cite{Hackbusch}), we get the
error-estimate
\begin{equation}
        \label{errest}
        \lambda^h_i-\lambda_i \leq C \lambda_i^{l+1} h^{2l},
\end{equation}
where $l$ is the polynomial degree of the used test functions.
This estimate shows that it is possible to compute an arbitrary
number of eigenvalues with a given accuracy by using a sufficiently
fine discretization.
Using standard piecewise linear functions, we obtain convergence of order $h^2$
so that small values of $h$ are required for high accuracy.
Accordingly, the dimension of matrices $A^h$ and $M^h$ becomes very
large, and solving the generalized eigenvalue problem (\ref{geneig}) is 
extremely costly.

Using test functions of higher degree yields a significantly
improved performance of the method.
We suggest to use a special modification of tensor product B-splines
with arbitrary coordinate degree. 
While, independent of $l$, B-splines may do with only one coefficient per
element, standard approaches of higher order require a much larger number, 
which is increasing with the degree. Since computation time to solve (\ref{geneig})
is growing rapidly with the number of degrees of freedom, this
aspect is crucial for achieving high accuracy.

So far, 
despite these advantages, B-splines have rarely been used in 
finite element applications, the main reasons being
Dirichlet boundary conditions and stability.  
A solution to these problems was proposed by H\"ollig,
Reif and Wipper \cite{HRWpaper}. 
Stability is achieved by linking potentially unstable B-splines near the boundary 
of the domain in an appropriate way to B-splines in the interior of the domain.
Following old, but little known ideas of Kantorovitch and Krylow,
homogeneous Dirichlet boundary conditions are ensured by multiplying 
all test functions by a weight function $w$ which vanishes on the boundary and is
positive inside the domain $\Omega$. 
It can be shown that the resulting weighted extended B-splines
(web-splines) form a stable basis and
possess the same approxima\-tion order as the underlying
tensor product splines space \cite{HRWpaper,HRWerror,Hbook}.

In this work we apply the WEB-spline method to approximate
eigenvalues of the Laplacian on the three-dimensional stadium
billiard $\Omega \subseteq \mathbb R^3$.
The implementation includes a specifically designed high accuracy integration 
algorithm. It is based on precomputed projections of the boundary grid cells
and iterated one-dimensional Gauss quadrature.

We use tensor product B-splines of degree $l=4$ on a grid with
$25 \times 25 \times 40$ cells. This leads to a generalized
eigenvalue-problem (\ref{geneig}) of dimension 9250.
To determine the eigenvalues, we use the Cholesky-decomposition
$M^h=LL^t$ to compute the matrix
\begin{equation}
        \label{simeig}
        B^h := L^{-1}A^h (L^t)^{-1}.
\end{equation}
The eigenvalues of $B^h$ are
just the generalized eigenvalues of $(A^h,M^h)$ so that
standard software
can be used to compute the $\lambda_i^h$.
Compared with iterative methods, the advantage of this procedure is
that one can be sure that no eigenvalues in the domain of interest are lost.
On the other hand, the maximal dimension that can be handled that way is
limited by the fact that $B^h$ is a dense matrix.

Of course, only a certain fraction of the eigenvalues $\lambda_i^h$ of $B^h$
provides a reasonable approximation of some $\lambda_i$.
The estimate (\ref{errest}) suggests that smaller eigenvalues are
approximated better than larger ones, but it does not provide
actual bounds since the constant $C$ is not known explicitly.
Computations on other domains, where the exact eigenvalues $\lambda_i$
are explicitly known, indicate that at least the first $16\%$ of the eigenvalues
$\lambda_i^h$, when ordered my modulus, have a relative error less than $0.1\%$
For the statistical evaluation in this paper, $1200$ out of $9250$ eigenvalues,
i.e., $13\%$ are used.

\section*{ACKNOWLEDGMENTS}

This work was supported in part by the DFG within SFB 634, the Centre of
Research Excellence Nuclear and Radiation Physics (TU Darmstadt), 
and by the U.S. Department of Energy under Contract
Nos. DE-FG02-96ER40963 (University of Tennessee) and DE-AC05-00OR22725
with UT-Battelle, LLC (Oak Ridge National Laboratory). B. D. thanks
the Joint Institute for Heavy Ion Research for financial support
during her stay at the Oak Ridge National Laboratory.  T.~P. thanks
the Institut f\"ur Kernphysik, Technische Universit\"at
Darmstadt, for financial support during his visits in Darmstadt.


\begin{thebibliography}{99}
%
%
\bibitem{V94}
J. J. M. Verbaarschot, 
\prl {\bf 72}, 2531 (1994).

\bibitem{VW00}
J. J. M. Verbaarschot and T. Wettig,
Ann. Rev. Nucl. Part. Sci. {\bf 50}, 343 (2000). 

\bibitem{Brody81} T. A. Brody, J. Flores, J. B. French, P. A. Mello,
A. Pandey, and S. S. M. Wong,
\rmp {\bf 53}, 385 (1981).

\bibitem{Friedrich}
H. Friedrich and D. Wintgen, 
Phys. Rep. {\bf 183}, 37 (1989).

\bibitem{Mirlin} 
A. D. Mirlin, 
Phys. Rep. {\bf 326}, 259 (2000).
%
\bibitem{Noeckel}
J. U. N\"ockel and A. D. Stone, 
Nature {\bf 385}, 45 (1997). 
%
\bibitem{Stock}
H.-J. St\"ockmann, J. Stein, Phys. Rev. Lett. {\bf 64}, 2215 (1990).
%
\bibitem{Sridhar}
S. Sridhar, Phys. Rev. Lett. {\bf 67}, 785 (1991).
%
\bibitem{R1}
H.-D. Gr\"af, H.L. Harney, H. Lengeler, C.H. Lewenkopf, C. Rangacharyulu, A. Richter, 
P. Schardt and H.A. Weidenm\"uller,
\prl {\bf 69}, 1296 (1992).
%
\bibitem{Weav}
R. L. Weaver,
J. Acoust. Soc. Am. {\bf 85}, 1005 (1989).
%
\bibitem{Elle}
C. Ellegaard, T. Guhr, K. Lindemann, H. Q. Lorensen, J. Nyg{\aa}rd, and M. Oxborrow,
\prl {\bf 75}, 1546 (1995).
%
\bibitem{QC}
H.-~J. St\"{o}ckmann, {\it Quantum Chaos - An Introduction}, (Cambridge 
University Press, Cambridge, 1999);
F.~Haake, {\it Quantum Signatures of Chaos}, 2nd edition, (Springer Verlag, 
Berlin 2001);
{\it Chaos and Quantum Physics}, edited by M.-J. Giannoni,
A. Voros, and J. Zinn-Justin (Elsevier, Amsterdam,1991).
%
\bibitem{GMW}
T. Guhr, A. M\"uller-Groeling, H. A. Weidenm\"uller, Phys. Rep. {\bf 299},
189 (1998).
%
\bibitem{Deus}
S. Deus, P. M. Koch, L. Sirko, \pre {\bf 52}, 1146 (1995).
%
\bibitem{Wirzba}
H. Alt, H.-D. Gr\"af, R. Hofferbert, C. Rangacharyulu, H. Rehfeld, A. Richter, P. Schardt, and A. Wirzba,
\pre {\bf54}, 2303 (1996).
%
\bibitem{SinaiExp}
H. Alt, C. Dembowski, H.-D. Gr\"af, R. Hofferbert, H. Rehfeld, A. Richter,R. Schuhmann, 
and T. Weiland,
\prl {\bf 79}, 1026 (1997).
%
\bibitem{Prosen}
T. Prosen, Phys. Lett. A {\bf 233}, 323 (1997), {\it ibid.} 332.
%
\bibitem{PS}
H. Primack and U. Smilansky,
\prl {\bf 74}, 4831 (1995); Phys. Rep. {\bf 327}, 1 (2000).
%
\bibitem{Sinai}
Y. G. Sinai, Russian Math. Surveys (2) {\bf 25}, 137 (1970).
%
\bibitem{Gutz}
M. C. Gutzwiller, 
J. Math. Phys. {\bf 11}, 1791 (1970); {\it ibid.} {\bf 12}, 343 (1971).
%
\bibitem{Gutzwiller}
M. C. Gutzwiller, {\it Chaos in Classical and Quantum Mechanics} (Springer,
New York, 1990).
%
\bibitem{Richt} C. Dembowski, B. Dietz, H.-D. Gr\"af, A. Heine,
T. Papenbrock, A. Richter, and C. Richter,
\prl{\bf 89}, 064101 (2002).
%
\bibitem{BaDu}
R. Balian and B. Duplantier,
Ann. Phys. {\bf 104}, 300 (1977).
%
\bibitem{PP}
T. Papenbrock and T. Prosen,
\prl {\bf 84}, 262 (2000).
%
\bibitem{Bunimovich}
L. A. Bunimovich, 
Commun. Math. Phys. {\bf 65}, 295 (1979).
%
\bibitem{Stadium3D}
T. Papenbrock,
\pre {\bf 61}, 4626 (2000).
%
\bibitem{Bun06}
L. A. Bunimovich and G. Del Magno,
Commun. Math. Phys. {\bf 262}, 17 (2006).
%
\bibitem{Taylor}
Y. Y. Bai, G. Hose, K. Stefa\'nski, and H. S. Taylor,
\pra {\bf 31}, 2821 (1985).
%
\bibitem{Sieber}
M. Sieber,
Nonlinearity {\bf 11}, 1607 (1998).

\bibitem{Sie}
M. Sieber, U. Smilansky, S. C. Creagh, and R. G. Littlejohn,
J. Phys. A {\bf 26}, 6217 (1993).

\bibitem{Moessner}
B. M\"o{\ss}ner,
{\it B-Splines als Finite Elemente}, Aachen (Shaker, 2006).
%
\bibitem{bgs}
M.~L.~Mehta, {\it Random Matrices}, 2nd ed. (Academic Press, San Diego, 1991).
%
\bibitem{Rosen}
N. Rosenzweig and C. E. Porter, Phys. Rev. {\bf 120}, 1698 (1960).
%
\bibitem{Leitner}
D. M. Leitner, Phys. Rev. E {\bf 48}, 2536 (1993).
%
\bibitem{Alt}
H. Alt, C.I. Barbosa, H.-D. Gr\"af, T. Guhr, H.L. Harney, R. Hofferbert, H. Rehfeld, and A. Richter,
\prl {\bf 81}, 4847 (1998).
%
\bibitem{SymBrech}
 B. Dietz, T. Guhr, H. L. Harney, and A. Richter, 
\prl {\bf 96}, 254101 (2006).
%
\bibitem{BGS}
G. Casati, F. Valz--Gris, and I. Guarneri, Lett. Nuovo
Cimento {\bf 28}, 279 (1980), M.V. Berry, Ann. Phys. (N.Y.) {\bf 131}, 163 (1981);
``Structures in semiclassical spectra: a question of scale'' in {\it The
Wave-Particle Dualism}, eds. S. Diner, D. Fargue, G. Lochak, and F. Selleri 
(D. Reidel, Dordrecht, 1984), 231;
O. Bohigas, M. J. Giannoni, and C. Schmit, Phys. Rev. Lett. {\bf 52}, 1 (1984).
%
\bibitem{heusler}
S. Heusler, S. M\"uller, A. Altland, P. Braun, and F. Haake,
\prl {\bf 98}, 044103 (2007).
%
\bibitem{Vergini}
E. Vergini and M. Saraceno, 
\pre {\bf 52}, 2204 (1995).
%
\bibitem{Hackbusch}
 W. Hackbusch,
 Theorie und Numerik elliptischer Differentialgleichungen,
 Stuttgart, 1996.
\bibitem{HRWpaper}
 K. H\"ollig, U. Reif, and J. Wipper:
 Weighted extended B-spline approximation of Dirichlet problems,
 SIAM J. Numer. Anal. 39:2 (2001), 442-462.
\bibitem{HRWerror}
 K. H\"ollig, U. Reif and J. Wipper:
 Error Estimates for the WEB-Method,
 Mathematical Methods for Curves and Surfaces: Oslo 2000,
 Vanderbilt University Press (2001), 195-209.
\bibitem{Hbook}
 K. H\"ollig:
 Finite Element Methods with B-Splines,
 Frontiers in Applied Mathematics 26, SIAM (2003).



%
\end{thebibliography}
\end{document}